\documentstyle[twocolumn,eqsecnum,aps]{revtex}
\def\be{\begin{equation}}
\def\ee{\end{equation}}
\def\bea{\begin{eqnarray}}
\def\eea{\end{eqnarray}}
\begin{document}
\draft
\title{Off-Forward Parton Distributions in 1+1 Dimensional QCD}
\author{Matthias Burkardt}
\address{Department of Physics\\
New Mexico State University\\
Las Cruces, NM 88003-0001\\U.S.A.}
\maketitle
\begin{abstract}
We use two-dimensional QCD as a toy laboratory to study 
off-forward parton distributions (OFPDs) in a covariant field 
theory. Exact expressions (to leading order in $1/N_C$) are 
presented for OFPDs in this model and are evaluated for some 
specific numerical examples. Special emphasis is put on 
comparing the $x>\zeta$ and $x<\zeta$ regimes as well as on 
analyzing the implications for the light-cone description of 
form factors.
\end{abstract}
\narrowtext
\section{Introduction}
Deeply virtual Compton scattering (DVCS) provides a novel tool 
to explore hadron structure. In contrast to deep-inelastic 
scattering (DIS), where one measures the imaginary part of the 
forward Compton amplitude only, DVCS allows measuring the 
off-forward Compton amplitude. From the parton point of view 
this implies that DVCS allows measuring off-forward matrix 
elements of parton correlation functions, i.e. on can access 
light-cone correlation functions of the form
\be
f_\zeta(x,t) \equiv \int \frac{dx^-}{4\pi} 
\langle p^\prime | \bar{\psi}(0) \gamma^+
\psi(x^-) |p\rangle e^{ixp^+x^-} , \label{eq:off}
\ee
where $x^\pm = x^0\pm x^3$ and $p^+=p^0+p^3$ refers to the usual 
light-cone components and $t\equiv q^2=(p-p^\prime )^2$ 
is the invariant momentum transfer. The ``off-forwardness'' 
(or skewedness) in Eq. (\ref{eq:off}) is defined to be 
$\zeta\equiv \frac{q^+}{p^+}$.\footnote{Note that we have 
chosen to follow here the conventions of Ref. \cite{ar}, which 
differ slightly from those in Ref. \cite{ji}.}

The main difference between these off-forward parton 
distributions (OFPDs) and the usual (forward) parton distribution functions 
is that $p^\prime \neq p$ in Eq. (\ref{eq:off}), i.e. since
the initial state is different from the final state in the above 
matrix element, OFPDs cannot be interpreted 
as parton densities. Although it is not yet entirely clear what intuitive physical information is contained in the off-forward
distributions, it is already clear that they play a `dual' role between
form-factors and parton distribution functions \cite{ji,ar}: 
when one integrates Eq. (\ref{eq:off}) over $x$ one obtains the 
matrix element of a local operator, i.e. form factors.
On the other hand, in the forward limit $\zeta=0$ and $t=0$,
one recovers the usual parton densities, and the case $\zeta=0$, $t\neq 0$
is related to impact parameter dependent parton distributions.
In a certain sense one can therefore interpret OFPDs as a
`parton decomposition of form factors', because $f_\zeta(x,t)$
gives the contribution of partons with momentum fraction $x$
to the form factor $F(t)$. 

Therefore, OFPDs might be of
tremendous help in studying the light-cone wave functions
of nucleons. For example, in the light-cone formalism, form
factors for $q^2<0$ can be evaluated in the Drell-Yan-West
frame where $q^+=0$. In this frame, the current is diagonal
in Fock space and form factors can be expressed in terms of 
overlap integrals between light-cone wave functions of the
target (see e.g. Ref. \cite{diehl}). 
Compared to the form factors, OFPDs give a much more
detailed information about light-cone wave functions since
they involve one less integration. 

One of the great 
difficulties in applying light-cone wave function based
phenomenology to form factors with $q^2>0$ is that one can
no longer go to a frame where $q^+=0$ and thus the current
operator is not necessarily diagonal in Fock space. As
a result, form factors for $q^2>0$ can no longer be expressed
as overlap integrals involving only the target's light-cone
wave function, but one must also include contributions where
the current first forms a meson which subsequently merges
with the target. OFPDs provide the unique opportunity to 
separate these contributions, since the region $x>\zeta$
involves only contributions where the current is diagonal in
Fock space, while $x<\zeta$ corresponds to the regime where
the current creates a $q\bar{q}$ pair.

Despite the similarities between ordinary parton distributions
and OFPDs in certain limiting cases, the physical 
interpretation of OFPDs in general is still less obvious. This 
fact makes it particularly difficult to model
OFPDs since it is not clear what properties one should expect
and therefore toy model studies may be very useful during
such a development stage. In such a situation, where one is 
studying a new observable with little known properties, it is 
often very useful to consider solvable toy models. 

For this purpose, we will use the  `t Hooft model \cite{thooft}
(1+1 dimensional QCD for $N_C\rightarrow \infty $) as an 
exactly solvable toy model to study OFPDs in a relativistic 
field theory. Despite being just a toy model, 
$QCD_{1+1}$ has a number of interesting features,
such as confinement, in common with $QCD_{3+1}$.
Of course, although toy models cannot predict what really 
happens in 3+1 dimensional QCD, they can
provide useful insight on typical features that one should
expect in a relativistic field theory: unlike most
phenomenological models, the `t Hooft model describes a
covariant field theory and its solutions still reflect 
the implicit constraints inherent to relativistic field 
theories, such as for example 
boost invariance and analyticity.

The paper is organized as follows. In Section II, form factors for 
the `t Hooft model are expressed in terms of light-cone wave functions.
because of the close connection between form factors and OFPDs, these
form the basis for deriving explicit expressions for OFPDs
in terms of light-cone wave functions in Section III. Explicit numerical
examples are presented in Section IV, and the results are 
summarized in Section V.

\section{Parton distribution functions and form factors in two-dimensional
QCD}
In the `t Hooft model \cite{thooft}, i.e. two dimensional QCD in the large 
$N_C$ limit, 
mesons are described as (non-interacting) bound states of
a quark-antiquark pair ($a\bar{b}$)
\be
M^2 \Psi(x) = \left( \frac{m_a^2}{x} + \frac{m_{\bar{b}}^2}{1-x} \right) \Psi (x)
+ \bar{g}^2 \int_0^1 \!\!\! dy \frac{ \Psi(x)-\Psi(y)}{\left(x-y\right)^2}.
\label{eq:thooft} 
\ee
Here $x$ denotes the momentum fraction carried by the quark in the 
infinite momentum (light-front) frame.
$\bar{g}^2 \equiv \frac{g^2N_C}{2\pi}$ is kept fixed in the 
$N_C \rightarrow \infty$ limit. 
The eigenvalues $M_n^2$ of Eq. (\ref{eq:thooft}) form a discrete spectrum 
and correspond to the invariant masses of meson resonances in this model.

Since the `t Hooft model is formulated in a light-front framework,
parton distribution, which have the physical meaning of momentum densities in 
the infinite momentum frame, can be easily calculated from the 
eigenfunctions to the
`t Hooft equation. One thus finds for the distribution function of
the quark $a$ in state with light-cone wave function $\Psi(x)$ 
\cite{einhorn}
\be
f_a(x) = \left|\Psi (x)\right|^2.
\label{eq:parton}
\ee
In Ref. \cite{einhorn}, one can also find exact expressions for the vector 
form factor. For details, the reader is referred to this very useful paper.
For example, for the matrix element of the $+$ component of the
vector current, which couples only to the quark, between two
meson states one finds
\be
\langle m, p^\prime | j^+(0) | n, p \rangle = 2P^+ F^+_{mn}(\zeta)
\ee
where
\bea
F^+_{mn}(\zeta) &=& \int_\zeta^1\!\!\!dx \Psi_m^*\!\left(\frac{x-\zeta}{1-
\zeta}\right) \Psi_n(x) \label{eq:einhorn}
\\
& & \!\!\!\!\!\!\!\!\!\!\!\!\!\!\!\!\!\!\!\!\!+\bar{g}^2 
\zeta^2\!\!\int_0^1 \!\!\!\!dv \!\!\int_\zeta^1 \!\!\!\!dy\ \Psi_m^*\!
\left(\frac{y-\zeta}{1-\zeta}\right)\!\frac{\Psi_n(\zeta v)-\Psi_n(y)}{(y-\zeta 
w)^2}
G\left(v;t\right).
\nonumber
\eea
and
\bea 
G(u;t)=\int_0^1\!\! dv\ G(u,v;t) \nonumber\\
G(u,v;t)=\sum_k \frac{\phi_k(u)\phi_k(v)}{t-\mu_k^2+i\varepsilon}
\label{eq:green}
\eea
is the `Green's function for an $a\bar{a}$ pair, i.e. the $\mu^2_n$ and $\phi_n$
in Eq. (\ref{eq:green}) are eigenvalues and eigenfunctions to the `t Hooft equation
with $m_{\bar{b}} = m_a$.

Here $q=p-p^\prime$, $\zeta = \frac{q^+}{p^+}$ and $t=q^2$.
In 1+1 dimensions, there is no transverse momentum and therefore the invariant
momentum transfer $t$ is related to the `off-forwardness' $\zeta$ through
energy conservation, i.e.
\be
M_n^2 = \frac{t}{\zeta} + \frac{M_m^2}{1-\zeta} .
\label{eq:target}
\ee 

In Section \ref{sec:off}, we will provide a interpretation of
OFPDs in terms of light-cone time ordered diagrams.
Since the Fock space interpretation of the various terms in 
form factors is very similar to the the terms contributing
to OFPDs, the reader is referred to Fig. \ref{fig:form}
for this purpose.

The first term on the r.h.s. in Eq. (\ref{eq:einhorn})
arises from pieces in the current operator which are diagonal
in Fock space. Those terms can be directly evaluated from
the overlap integral between the light-cone momentum space
wave functions in the two particle sector.
The second term on the r.h.s. in Eq. (\ref{eq:einhorn})
corresponds to pair creation terms in the current operator,
i.e. physically it corresponds to the emission of a virtual
meson which is subsequently absorbed by the external current.
Naively, one may not have expected that such a contribution 
survives the $N_C\rightarrow \infty$ limit where meson-meson
vertices are suppressed by one power of $\frac{1}{\sqrt{N_C}}$.
However, the vacuum to meson matrix element of the current
operator scales like $\sqrt{N_C}$, which compensates for the
$\frac{1}{\sqrt{N_C}}$ suppression of the triple meson vertex.

The Green's function $G(u,v;t)$, in the second term on the 
r.h.s. of Eq. (\ref{eq:einhorn}), describes the 
non-perturbative interaction of the $q\bar{q}$ pair emanating 
from the external current before it interacts with the target 
meson. Following Ref. \cite{einhorn}, we describe this 
interaction by inserting a complete set of meson states.
It should be emphasized that Eq. (\ref{eq:einhorn}) is
exact to leading order in $\frac{1}{N_C}$, i.e. it should
have all properties that one would expect from a form factor
in a relativistic field theory.
For further details, as well as a derivation of Eq. 
(\ref{eq:einhorn}), the reader is referred to Ref. 
\cite{einhorn}. 

\section{Off forward distribution functions}
\label{sec:off}
\begin{figure}
\setlength{\unitlength}{1cm}
\begin{picture}(6,7.5)(3.5,12.2)
\includegraphics{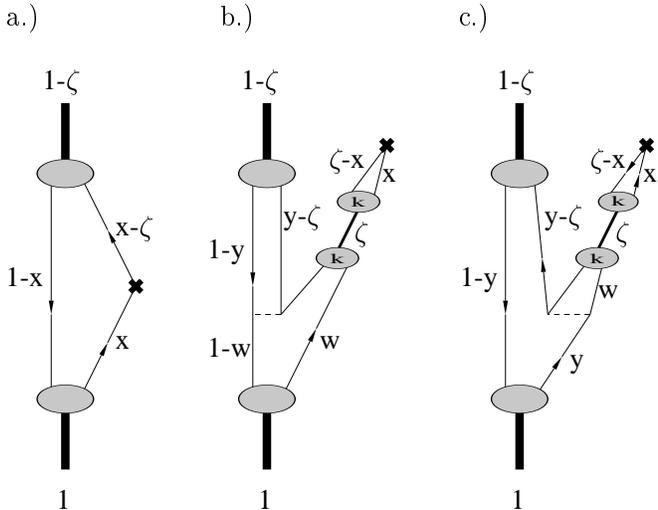}
\end{picture}
\caption{Light-cone time-ordered diagrams, illustrating
the physics of the various contributions to 
$f_\zeta(x)$. 
a.) Diagrams diagonal in Fock space contributing for
$x>\zeta$.
b.),c.) diagrams off-diagonal in Fock space, which are relevant
for $x<\zeta$.
All momenta are in units of $p^+$. The larger gray blobs
represent the (light-cone) wave function of the incoming
and outgoing meson respectively.
The two smaller blobs connected by a meson propagator in
b.) and c.) represent the insertion of a complete set of 
meson states.}
\label{fig:form}
\end{figure}

Since Einhorn's result for the form factors (\ref{eq:einhorn}) is already
expressed in terms of light-cone variables, it is a straightforward
exercise to isolate the off forward distribution functions.
For $x>\zeta$, only the piece diagonal in Fock space contributes and one
finds (Fig. \ref{fig:form}a)
\be
f_\zeta(x) = \Psi^*\!\left(\frac{x-\zeta}{1-\zeta}\right) \Psi(x) 
\quad \quad \quad \quad \quad (x>\zeta).
\label{eq:diag}
\ee
Note that Eq. (\ref{eq:diag}) saturates the positivity
constraint for OFPDs \cite{soffer}.
Eq. (\ref{eq:diag}) reduces to the usual parton distribution [Eq. (\ref{eq:parton})]
for $\zeta=0$, i.e.
\be
f_{\zeta=0}(x) = f(x).
\ee
For $x<\zeta$, the bilocal current in Eq. (\ref{eq:off})
creates a $q\bar{q}$ pair with momentum fraction $\zeta$, i.e.
only the off diagonal piece contributes and one finds
(Fig. \ref{fig:form}b,c)
\be
f_\zeta (x) = f_\zeta^b (x) + f_\zeta^c(x)
\quad \quad \quad \quad \quad (x<\zeta),
\label{eq:offdi}
\ee
where
\be
f_\zeta^b(x)=  
\bar{g}^2 \!\!
\int_0^\zeta \!\!\!\!dw \!\!\int_\zeta^1 \!\!\!\!dy\ \Psi^*\!
\left(\frac{y-\zeta}{1-\zeta}\right)\!\frac{\Psi(w)}{(y-w)^2}
G\left(\frac{w}{\zeta},\frac{x}{\zeta};t\right)
\ee
and
\be
f_\zeta^c(x)= -  
\bar{g}^2 \!\!
\int_0^\zeta \!\!\!\!dw \!\!\int_\zeta^1 \!\!\!\!dy\ \Psi^*\!
\left(\frac{y-\zeta}{1-\zeta}\right)\!\frac{\Psi(y)}{(y-w)^2}
G\left(\frac{w}{\zeta},\frac{x}{\zeta};t\right).
\ee
Like Eq. (\ref{eq:einhorn}), Eqs. (\ref{eq:diag}) and
(\ref{eq:offdi}) are exact to leading order in $\frac{1}{N_C}$!
The physical interpretation of Eq. (\ref{eq:offdi}) is as
follows: for $x<\zeta$, the current in Eq. (\ref{eq:off})
creates a quark (with momentum $xp^+$ and an antiquark
with momentum $(\zeta-x)p^+$. This $q\bar{q}$ pair can of
course interact with itself before merging with the external 
meson states. Similar to the expression for the form factor,
this interaction among the $q\bar{q}$ pair can be described 
by the Green's function $G(u,v;t)$, which suggests the same
physical picture of a $q\bar{q}$ pair, which forms a meson
before being absorbed by the target meson. 

\section{Numerical results}
In the numerical calculations, we used basis functions of the form
\be
\chi_n(x) = x^\beta (1-x)^\beta P_n(x),
\ee
where $\beta$ is determined from the boundary condition
$\pi \beta \cot \pi \beta = 1-\frac{m^2}{\bar{g}^2}$
and $P_n(x)$ is a polynomial of $n-th$ order.
We found that $n\leq 10$ basis function are sufficient to achieve numerical 
convergence (first five figures of ground state masses stable and
OFPDs did not show any visible dependence on the size of the basis).

The `t Hooft equation (\ref{eq:thooft}) contains two dimensionful parameters:
the gauge coupling $\bar{g}$ and the quark mass $m$. Of course, since the 
`t Hooft model is only a toy model, there is no fundamental reason to
associate any parameters in the model with experimental numbers in the real 
world. However, since we are trying to gain an intuitive understanding in
general for regimes where quark masses are very small to `medium sized'
(u,d and s quarks), we adopt the following procedure to fix the parameters:
First the coupling $\bar{g}$ is adjusted to yield the physical string tension
in QCD, i.e. we demand
\be
\frac{\pi}{2}\bar{g}^2 \stackrel{!}{=} 0.18 \left(GeV\right)^2 ,
\ee
yielding $\bar{g}=340 MeV$.

The quark masses are determined by demanding that the masses
of ground state $q\bar{q}$ and $q\bar{s}$ mesons coincide with 
the masses of $\pi$ and $K$ mesons respectively,
yielding
\bea
m_q&=& .045 \bar{g}\nonumber\\
m_s&=& 1.1 \bar{g} .
\eea

\begin{figure}
\setlength{\unitlength}{1cm}
\begin{picture}(6,18)(4.8,7)
\includegraphics{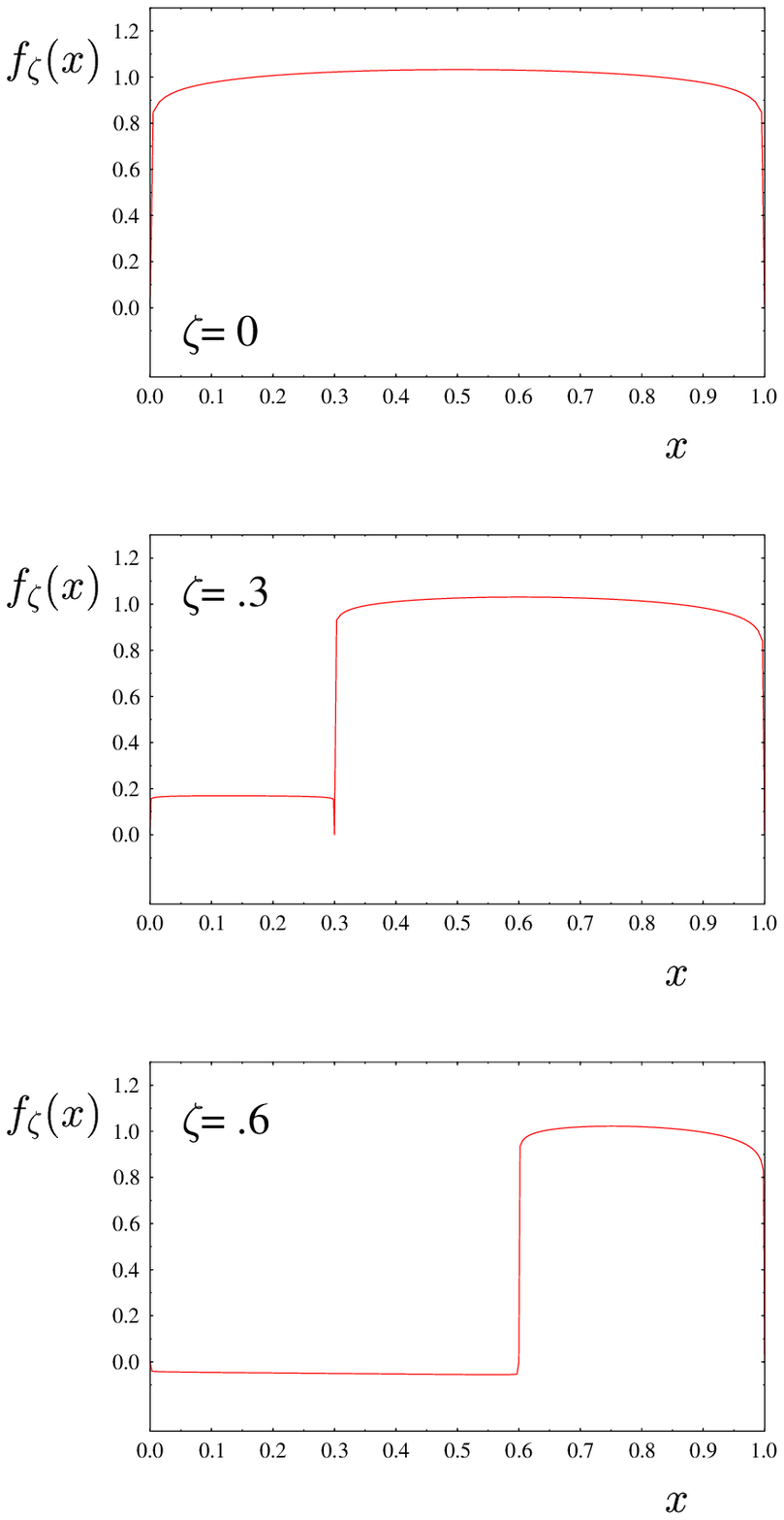}
\end{picture}
\caption{Off-forward parton distributions in a $\bar{q}q$ meson with
light quarks ($m_\pi=140 MeV$) for $\zeta=0,\, 0.3$ and $0.6$.
The $\zeta=0$ result is the conventional parton distribution. 
}
\label{fig:fzeta}
\end{figure}

\begin{figure}
\setlength{\unitlength}{1cm}
\begin{picture}(6,18)(4.8,7)
\includegraphics{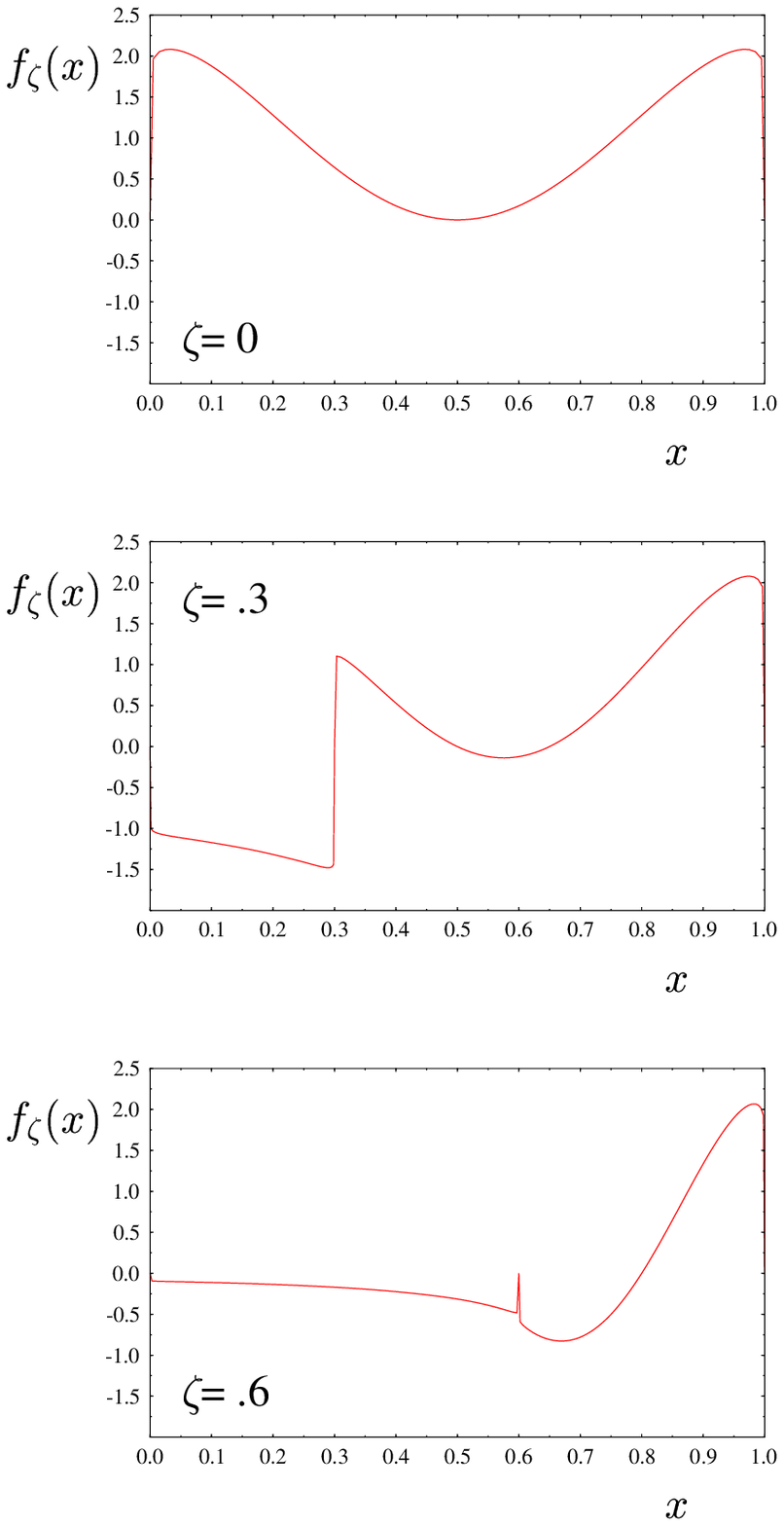}
\end{picture}
\caption{Same as Fig. \ref{fig:fzeta}, but for the first orbitally
excited meson state.}
\label{fig:fzeta2}
\end{figure}

\begin{figure}
\setlength{\unitlength}{1cm}
\begin{picture}(6,18)(4.8,7)
\includegraphics{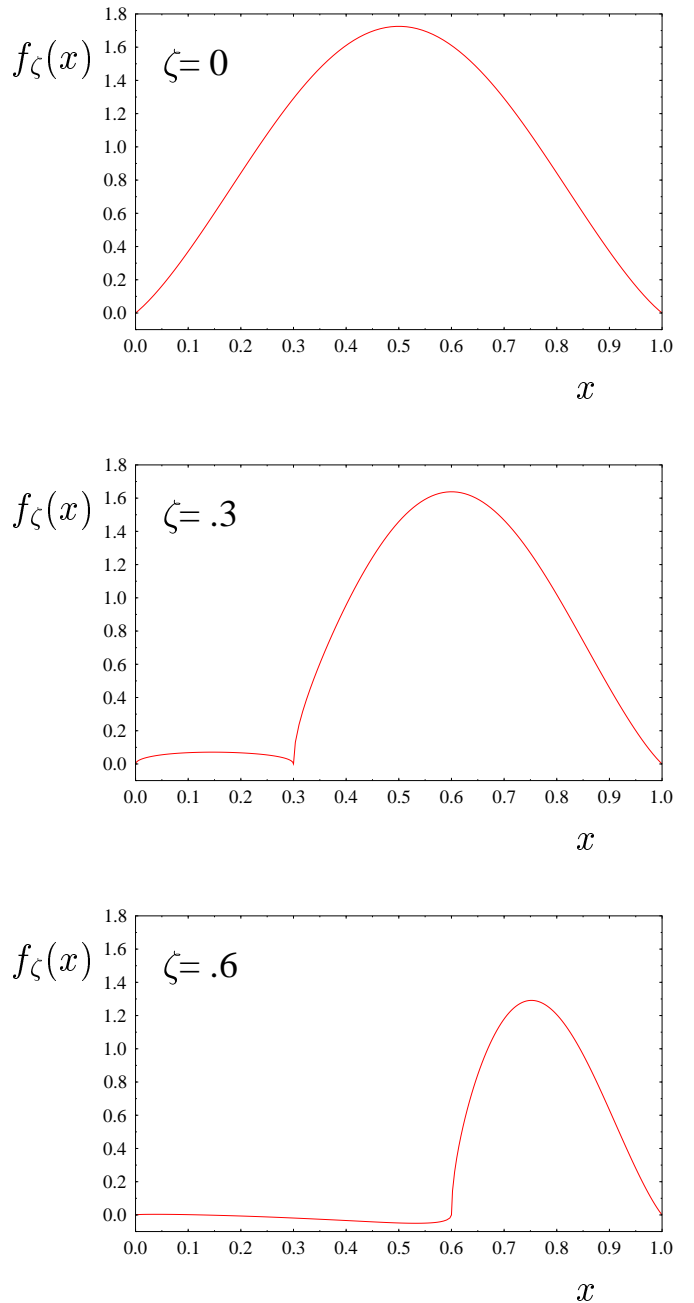}
\end{picture}
\caption{Same as Fig. \ref{fig:fzeta} but for a quark mass on the
order of the strange quark mass.}
\label{fig:fzetas}
\end{figure}

Numerical results for the OFPD $f_\zeta(x)$ for $\bar{q}q$ and $\bar{s}s$
ground state mesons are shown in Figs. \ref{fig:fzeta}
and \ref{fig:fzetas} respectively.
Results for the first excited meson state are shown in
Fig. \ref{fig:fzeta2}.
For the $\pi$ meson in the `t Hooft model the distribution amplitude
is nearly flat. As a result, the parton distribution function
($\zeta=0$ in Fig. \ref{fig:fzeta}), which is just the square of the
wave function in this model, is also nearly flat. For nonzero $\zeta$,
Fig. \ref{fig:fzeta} exhibits several interesting features.
First of all, $f_\zeta(1)=f_\zeta(\zeta)=f_\zeta(0)=0$. Secondly,
$f_\zeta(x)$ is nearly flat for $0<\zeta<x$ as well as for 
$\zeta<x<1$. 

In the `t Hooft model, the (target-) meson 
wave function is nearly flat (for small $m_q$) and vanishes (sharply)
at the boundary. Since $f_\zeta(x)$ is just the overlap of the wave function 
with itself, these features are reflected in $f_\zeta(x)$ for 
$\zeta < x < 1$, i.e. rapid rise from zero near $x=\zeta$, nearly flat
for $\zeta < x < 1$ and a rapid drop near $x=1$. In 3+1 dimensional QCD
one would also expect that $f_\zeta(1)=0$ because parton distributions
vanish at that point as well. However, since parton distributions in
$QCD_{3+1}$ do not 
vanish at $x=0$, one would not expect $f_\zeta(\zeta)$
to vanish either.
The physics of the regime $0<x<\zeta$ is entirely different. Here the
$x$ dependence comes from the $x$ dependence of the distribution
amplitude of the meson that couples to the external current. In general,
that contribution is complicated since there is an infinite sum of mesons
contributing. However, when $t$ is near a pole in this sum, that particular
meson will of course
dominate and as a result the $x$ dependence will be proportional
to the distribution amplitude of that particular meson. This is the case 
for small quark masses in $QCD_{1+1}$! In two dimensions, the `$\pi$'-meson
couples not only to pseudoscalar  but also to vector currents and thus
the `$\pi$' does contribute to the sum over meson poles in the Green's function
relevant for vector currents  (\ref{eq:green}). 

In two dimensions, $t$ and $\zeta$ are not independent variables, but are
related to each other (and the target mass) through Eq. (\ref{eq:target}).
Since the pion mass
goes to zero for small $m_q$ and since $t$ is proportional to the target
mass$^2$ (here also the $\pi$) we are in a situation where the $\pi$ pole
contribution dominates in the sum over meson poles. As a result, the
$x$ dependence for $0<x<\zeta$ is, to a good approximation, proportional to
$\Psi_\pi(\frac{x}{\zeta})$. 
The coefficient of proportionality depends on $\zeta$ in a 
complicated way [described by Eq. (\ref{eq:offdi})].
In fact it even changes sign around $\zeta\approx 0.5$ since there are two
competing terms with opposite sign.\footnote{As is illustrated 
in Fig. \ref{fig:form} b. and c., the gluon from the
annihilation vertex can be absorbed either by the quark or
the antiquark!}

This close connection between OFPD for $0<x<\zeta$ and
distribution amplitudes as for example been pointed out already
in Ref. \cite{ar} where it was noticed that OFPD evolve
with the Brodsky-Lepage kernel for distribution amplitudes
when $0<x<\zeta$, i.e. the UV part of OFPD and distribution
amplitudes are the same in this regime. 

In fact, one can use this connection to measure the light-cone
distribution amplitudes of mesons contributing in the sums over meson 
poles:
the $3+1$ dimensional generalization of Eq. (\ref{eq:offdi}) reads
\be
f_\zeta(x,t) = \sum_n \frac{c_n(\zeta,t)}{t-M_n^2} 
\phi_n\left(\frac{x}{\zeta}\right),
\ee
where $\phi_n(z)$ is the $n$-th meson's distribution amplitude,
and $c_n(\zeta,t)$ are some coefficients characterizing the
coupling of the mesons to the target. 
It is therefore conceivable that, by extrapolating in $t$ to the lowest
meson pole, to use OFPD in the
regime $0<x<\zeta$ as a tool to measure meson distribution
amplitudes (e.g. the $\rho$-meson distribution amplitude,
which is otherwise hard to access)

Although $f_\zeta(x)$ in $QCD_{1+1}$ is continuous and vanishes
for $x=0,\zeta$ and $1$, these properties are not very clearly
visible in Fig. \ref{fig:fzeta} due to the rapid rise of
the `t Hooft wave function near the endpoints for small $m_q$.
These features are better illustrated for heavier quarks,
where the wave function rises less rapidly near the endpoints.
Results for `strange' quarks 
\footnote{See the discussion above on how the parameters in 
the `t Hooft equation were fixed.} are shown in Fig. 
\ref{fig:fzetas}.

Up to trivial kinematic factors, $\int_0^1 dx f_\zeta(x)$ yields 
the form factor. For a ground state meson target, even for 
small quark masses, and for typical values of $\zeta$, 
Fig. \ref{fig:fzeta} shows that
most of the contribution to this integral arises
\noindent 
from the region 
$x>\zeta$, i.e. from the terms that are diagonal in Fock space
(\ref{eq:diag}). This is a very surprising result since it
should cost only very little energy to create a $q\bar{q}$
fluctuation which could then couple to the target.
\footnote{Note that the lightest meson in $QCD_{1+1}$, which
becomes massless as $m_q\rightarrow 0$, {\it does} contribute
to the sum over meson poles in Eq. (\ref{eq:offdi}).}
Numerically, there is of course a large cancellation between
the two terms in Eq. (\ref{eq:offdi}). These two terms 
correspond to the two possibilities that the 
instantaneous `gluon', which emerges when the antiquark from 
the current gets absorbed, couples to the quark or antiquark
in the target meson. Because these two terms almost seem to
cancel each other, it is suggestive to interpret this result 
in terms of a `screening phenomenon', indicating a `small size' 
\footnote{It is not clear which definition of `size' really
sets the scale here, which is why we were unable to quantify
this argument any further.}
for the ground state meson in the chiral limit. 
However, beyond this very handwaving picture,
we have no intuitive explanation for this surprising dynamical
suppression of the off-diagonal terms in the form factor for
the ground state meson in this model.

\section{Summary}
We have presented exact expressions for off forward parton distributions
in the 't Hooft model ($QCD_{1+1}$ with $N_C\rightarrow \infty$).
These expressions, together with their numerical evaluation,
illustrate a number of features that should be generic for OFPDs in
any relativistic field theory (regardless of the number of dimensions).

Comparing light-front Fock space overlap integrals for form factors 
(\ref{eq:einhorn})
and OFPDs (\ref{eq:diag},\ref{eq:offdi}) \footnote{For $\zeta>x$ 
the corresponding expressions in $QCD_{3+1}$ can for example be found in
Ref.\cite{diehl}.}
, it is evident that OFPDs can be interpreted as `parton 
decompositions of form 
factors'. In fact, the only difference between form factors and OFPDs is
the fact that in OFPDs the momentum of the `probed quark' is not integrated
over but rather kept fixed at momentum fraction x.
Of course, in the `t Hooft model, where the wave functions depend only
on one parameter, and therefore the Fock-space diagonal contribution
to the form factor involves only one integration, this means that the OFPD
is given by a mere product of the wave function with itself. 
This illustrates that, particularly for large $x$, where contributions 
from higher Fock components should be suppressed also in 3+1 dimensions,
OFPDs provide much more direct information on the light-cone wave function
of the target than form factors.

While the $x$ dependence for $x>\zeta$ depends only on properties of the 
target hadron's light-cone Fock space wave functions, this is no longer the 
case for $x<\zeta$. In the latter regime, the $x$ dependence is governed by
the light-cone wave functions of mesons coupling to the external current.
In fact, Eq. (\ref{eq:offdi}) reflects that for a fixed value of $\zeta$,
each meson contributes with some $\zeta$-dependent coefficient times its
light-cone wave function. Although it will in general be difficult to
disentangle this sum over meson poles into contributions from individual 
mesons, the lightest state with the appropriate quantum numbers should
give a dominant contribution for small $t$. For example, for unpolarized
OFPDs, where the $\gamma$-matrix structure of the probe is that of a 
vector current, one would be most sensitive to contributions from $\rho$
and $\omega$ mesons and the success of the vector meson dominance picture
for form factors suggests that there should be little contamination 
from higher states at small $t$. In the case of polarized OFPDs, where
the $\gamma$-matrix structure is that of an axial vector current, the
dominance of the lowest pole (the $\pi$) should be even more pronounced
at small $t$.

In the `t Hooft model we found that $f_\zeta(x)$ vanishes for $x=\zeta$
and is continuous at this point. The fact that $f_\zeta(x)$ vanishes for
$x\rightarrow \zeta$ from below comes from the fact that meson 
distribution amplitudes vanish at the end-points in this model and,
at least if distribution amplitudes are close to their asymptotic form in 
$QCD_{3+1}$, one would expect the same behavior there as well.
The situation is different for  $x\rightarrow \zeta$ from above, since there
one would expect significant contributions from higher Fock components.
In fact, $f_\zeta(x)$ might even diverge at this point.

In the introduction, we pointed out that OFPDs allow to
decompose form factors into their dependence on the
light-cone momentum of the active parton.
It is very remarkable that, despite the fact that a vector
meson dominance description for the form factor works
extremely well for small $m_q$ \cite{mende}, only a very
small portion of the form factor for a ground state meson
comes from the $x<\zeta$ region. Instead, for typical
values of $\zeta$, most of the contributions to the
form factor arise from $x>\zeta$, where the current is
diagonal in Fock space and the form factor can be expressed
as an overlap between light-cone wave functions. It would
be very interesting to study if a similar suppression of
the $x<\zeta$ contributions also takes place in $QCD_{3+1}$.

\noindent {\bf Acknowledgments}
This work was supported by a grant from DOE
(DE-FG03-95ER40965) and by TJNAF.

\end{document}